# Information Centric Networking based Handover Support for QoS Maintenance in Cooperative Heterogeneous Wireless Networks

Thesis Supervisor: Dr. Éric Renault

Thesis Advisors: Dr. Djamal Zeghlache

Submitted By : Muhammad Shoaib Saleem

Wireless Networks and Multimedia Services Department (RS2M)

Version 1

August 15, 2011

**Abstract**

Network of Information (NetInf) is a term coined for networks which unlike contemporary network are not node centric. As the name indicates, information supersedes nodes in the network. In this report, we propose an architecture of mobile node for NetInf. We call it NetInf Mobile Node. It is an extension of the basic node architecture proposed for NetInf. It is compatible to NetInf and TCP/IP based networks. The Virtual Node Layer modules in the architecture provide support for managing mobility, power consumption of the node as well data relaying/storing services. Inner/Outer Locator Construction Routers (I/O LCTR) are two functions introduced in NetInf mobile nodes to operate between NetInf and non-NetInf sites. The basic purpose of NetInf mobile node is to maintain the QoS during mobility events. The handoff/handover are critical situations during mobility where chances of QoS degradation of an ongoing session are high. This report presents one such scenario in which QoS of an application is maintained during a handoff situations in heterogeneous wireless network environment through our proposed algorithm.

# Contents

# List of Figures

# 1 Introduction

The genesis of NetInf is influenced by legacy and innovative networking technologies. The features which NetInf exhibits is the melange of existing and innovative solutions. Among them is the increased use of the overlay networks in the contemporary Internet. The P2P networks for file sharing are very popular (e.g. E-mule, Skype, Bit Torrent etc). Since these overlays are distributed networks, they share and take away the load from the central server. NetInf intends to incorporate these functions in its architecture. The notion of information in NetInf is bigger than what it is today considered in the Internet. NetInf considers not only virtual entities as information objets (like web pages, text files etc) but it also recognizes real world objects as information and part of its network. [1] discusses different research scenarios for NetInf. NetInf introduces terms like Information Objects(IOs), Data objects(DOs) and bit level objects(BOs) for information description in the network. IOs represent semantics of the object, DOs are the reference/locator to the real bit patterns i.e. BOs. The representation of real world objects in NetInf can be made possible by presenting them as virtual entities and integrating them in the NetInf.

In order to make information centric networking competetive enough to replace the legacy TCP/IP protocol stack, it should provide solutions for the existing problem in All-IP networks. The new architecture must be backward compatible with the existing Internet architecture so that the replacement can be done in an incremental way which is easy as well as economically feasible. The range of issues that contemporary IP-networks face are many folds. As the current Internet was not designed to cater every problem, hence, most of the solutions today are add-on. With the advancement in mobile telephony, new protocols were developed to handle mobility. The performance of these protocols remained steady in early years. However, in the last decade, the overwhelming development of new applications accessable on wireless networks through smart devices has urged to have more efficient algorithms and protocols. There are two possible solutions. Either, as usual, provide patch up solutions or design a clean slate architecture which provides built in solutions to all the problems and issues exhibited by TCP/IP based network architecture.

In couple of years, desktop Internet users will be outnumbered by mobile Internet users. One demand from user perspective is to maintain the QoS during mobility. In a heterogeneous network enviroment this requirement becomes inevitable, mostly in vertical handover situations where there is a break before make situation occurs whenever the service/application connectivity requirements change. In information centric networking, aspects like, support for mobility management is considered to be the part of the architecture. There have been many research projects in recent years to develop information/content/data oriented networks. For example, 4Ward [2], PSIPR [3], SAIL [4], CCN [5] and many others are the projects which have started to formulate and design an architecture that provides solutions to existing problems faced by the IP-networks of today. The common point in all these projects is to develop a framework for an information centric network. The approach can be different but the cornerstone of their proposed architecture is to have a network which is information centric. Each project has proposed differnet methods and procedures for supporting mobility, multihoming, naming and addressing, security, data dissemination and storage, routing and forwarding, QoS and QoE etc.

QoS is closely related to mobility management in wireless networks. To maintain QoS of an application during an ongoing session is mostly disrupted during handover/handoff scenarios. Today, heterogeneous wireless network environment demands seamless handover between various network access technologies. A lot of work has been done so far in this domain. Today, factors that trigger handover are not jsut limited to the measured value of the Received Signal Strength (RSS). The list involves data rate requirement for a particular application, end user demand for high/low data rate, packet loss and more. There is a wide range of protocols and algorithms across all the layers in a TCP/IP protocol stack for supporting mobility. For example, Mobile IPv4/IPv6 and LIN6 in Network layer. The Stream Control Transmission Protocol (SCTP) and the Datagram Congestion Control Protocol (DCCP) provide mobility support at trasport layer level. Session Initiation Protocol (SIP) , Dynamic DNS (DDNS) and MOBIKE are few examples of application layer protocols. The performance of these protocols is limited due to lack of cross layer cooperation between the lower and higher layers. The dual nature of an IP address, acting as a locator as well as an identifier degrades the efficiency of these protocols during mobility event. The separation of locator from its identifier is to counter this issue. Host Identity protocol (HIP), Locator Identifier Separation Protocol (LISP) are example of such proposals. HIP works in collaboration of transport and network layer. LISP is a network based approach and focusses on limiting the size of routing tables and improving scalability and routing system. LISP extention, LISP Mobile Node has multiple design goals including wide range of communication possibilities in different mobility cases.

The use of the devices capable of accessing different radio interfaces urges to have an architecture that facilitates seamless handover of sessions among different wireless acces technologies. IEEE 802.21-Media Independent Handover (MIH) standard supports collaboration between various access link layer access technologies. The MIH consists of a framework for providing services to its users. The framework includes various entities for transmitting and receiving messages to share information about various access networks's capacity. In other words, this framework defines a protocol stack, implemented on each mobile device for seamless handoff. The cross layer cooperation has also been studied extensively [6],[7],[8],[9]. The information exchange between different layers of the stack improves the overall handover process by avoiding false alarm signals and minimizing the latency during handover to reconnect or update.

There is a history of extensive work of mobility management. However, even after so much effort, the domain is still alive and very active. The reasons are quite obvious. At each layer of the stack, the requirements for mobility management demands different approaches (as mentioned earlier). Moreover, rapid advancement in mobile communication encourages to develop new ideas and frameworks. Ubiquitous QoS support in the wireless network environment is a big challenge. In urban areas, where problems like data traffic congestion, channel fading, and interference result into intolerable disconnectivity, poor coverage and lack of required QoS. We introduce the virtual node concept in out study. In our proposed virtual node framework, we introduce :

- A virtual node layer (VNL) in the NetInf MN. This VNL is a programming abstraction. The concept has been used before [10],[11],[12] but in the context of information centric networking, the idea is novel.

- We introduce a central entity known as Central Control Unit (CCU) which supports VNL coordination between various wireless network technologies specially during handover scenarios. CCU also records and updates the mobility pattern of the mobile nodes, predicting mobile node motion and allocation of mobility zones for virtual nodes.

- Our proposal emphasizes collaboration of network and the end user. The mobility events discussed are not entirely network controlled neither mobile controlled. The VNL with its modules, explained later in this report, along with cross layer cooperation, supports mobility by making it smooth and seamless.

The later sections are organized as follows. In section 2, mobility management in Network of Information, NetInf Mobile Node architecture is discussed along with VNL detailed description and cross layer support during handover. Section 3 presents VNL working principle explained through a scenario and the algorithm proposed. Section 4 briefs over the extended function of NetIng Mobile Node and is followed by conclusion and future work.

## 2  Mobility Management in NetInf

To handle mobility issues in NetInf, integrated name resolution and routing schemes are proposed. The proposed solutions include Multiple Distributed Hash Table (MDHT) [13] approach for core network, Late Locator construction (LLC) [14] scheme, which is an extention of MDHT, for access networks and autonomous local resoultion using multicast (providing access to all local content even when the network is not connected to core network). In MDHT, as the name indicates, DHTs are arranged in a hierarchical manner. As an example , in an Internet Service Provider (ISP), there are four levels,

- Access node level
- Point of Presence
- Autonomous System level
- Global Internet level

However, the structure can be changed depending upon the size of the network. There are name resolution platforms/nodes at each level of MDHT. These special platforms/nodes are addressed as Dictionary Nodes (DN). They are actually network based implementation of a DHT system. They perform name resolution and location look up service both in local and global scope of the network. The object (information) is publshed and duplicated at all levels. As far as LLC is concerned, it separates core network routing from edge edge network routing. It uses path based locator. In LLC, packets are forwarded in a connectionless manner. Path based locator for a object consists of a core edge router prefix appeneded with a sequence of identifiers that describes an internetwork path across a sequence of edge networks towards the host, hosting destination object. In general, LLC employs an object identifier/locator split mechanism. The common ground between MDHT and LLC schemes is that any one can be switched in depending upon requirement. It means what kind of mobility case NetInf is dealing with. In case of mobile terminal mobility, LLC is taken into account and for network mobility, MDHT scheme is approached.

The evaluation of both schemes namely, MDHTs and LLC, encourages the need to have more optimization in terms of reducing the handover latency during the mobility event. Further, in case of intermittent connectivity the delays are intolerable. One research challenge also indicates to have an interaction betweel mobility and caching to reduce the over all look up delay. This latency issue is mostly concerened with the mobile access networks that are not part of the actual core netwrok. These mobile access networks include all radio networks ranging from mobile telephony (3G, 4G, GSM, CDMA etc) to WiFi and WiMax.

The Issues discussed above are very crucial in multiaccess enviroment. In a large metropolitan environment, multitechnology interfaced mobiles device demand always best connected connection for better QoS experience. For real time streaming, smooth shift or handover between different access technologies is anticipated. Factors mentioned earlier like congestion, channel fading and unprecedented mobility of users are the challenges hinder good service. In the next subsection, we present our proposal of a mobile node architecture for information centric networks which is also backward compatible with the comtemporary TCP/IP based nodes centric network.

## 2.1 NetInf MN Architecture

Mobility management in a wireless environment has always been an important issue. There are many solutions proposed to target the problems faced by mobile nodes during mobility. The current Internet architecture is TCP/IP based and mobile nodes use Mobile IP or IP mobility, which is a protocol developed for mobile devices, to roam from network to network maintaining their permanent IP addresses. However, with the inception of the idea of a network of interconnected things, generally termed as Internet of Things, motivated to have a new Internet architecture, where information supersedes the node centric concept of networking.

In NetInf, the node architecture defines the general framework for a typical information centric network node. Based on this architecture, we defined Network of Information Mobile Node (NetInf MN) [15]. Along with new features introduced, NetInf Mobile Node is compatible with any environment or wireless access technology. There are some similar characteristic between NetInf MN and Locator Identifier Separation Protocol Mobile Node (LISP MN) [16]. However, LISP Mobile Node has a totally different working framework. It includes servers that facilitate mapping from end point identifiers to locators. NetInf has totally different infrastructure for name resolution. Still there are some common features that NetInf MN shares with LISP MN like both nodes encourages to :

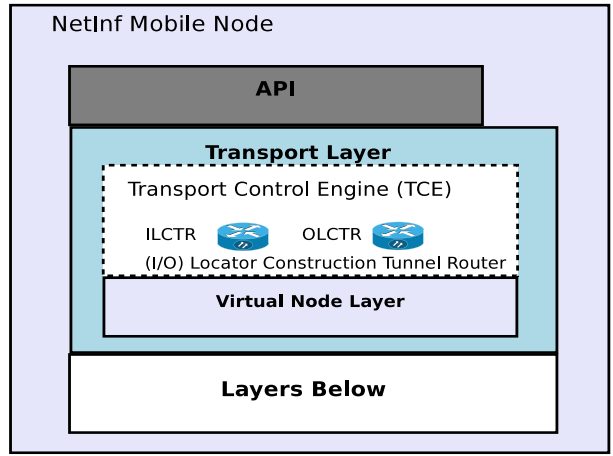

Figure 1: Network of Information Mobile Node

- Keep ongoing sessions alive during mobility even if all nodes are moving simultaneous.

- Have the possibility of simultaneous connections.

- Make mobile node to act as a server.

In this work, the proposed solution for mobility management in network is presented in the form of a mobile node architecture. Fig. 1 presents the NetInf MN architecture. We have an application layer which provides services to users. This API gets services from the transport layer below it. Within the tranport layer we have Tranport Control Engine (TCE). Generally in NetInf, TCE is responsible for the coordination of protocols used for accessing NetInf objects. In our design, we propose to include Inner Locator construction tunnel Router (ILCTR) and Outer Locator Construction Tunnel Router (OLCTR) functionalities in TCE. Virtual Node Layer (VNL), ILCTR (Inner locator Construction Router) and OLCTR (Outer Locator Construction Router) within the transport layer are responsible for the mobility management of NetInf MN. The working of these routing functions is explained in later section.

The work presented here highlights the features of the NetInf MN architecture and VNL and is complemented with its working principle, advantages it exhibits and its supporting role within the network.

### 2.1.1 Virtual Node Layer (VNL)

VNL is a programming abstraction in NetInf-MN architecture. Our proposed scheme involves cross layer (Layer 2 & 3) cooperation with VNL to support handoff. Link and network layer parameters (i.e. mobile terminal speed and handoff delay signal respectively) influence mobility management. It is proposed here that in a heterogenous environment, a Centralized Control Unit (CCU) Fig.2 exists that assists on behalf of network for managing mobility. Since cellular networks and WiMax covers large area, we propose that each base station is equipped with a CCU.The VNL unit Fig.3, together with the Link and Network layer, estimates the above parameters (to initiate handoff ) and update the CCU. It provides following basic funtions.

- VNL together with the Layer (2+3), assists (inter/intra) domain handovers.

- In a non active state or with minimum battery life left, power management by lowering the paging signals for location update by mobile terminals.

- NetInf-MN provides temporary storage or relay service to store the data packets in challeneged enviroment where the network connectivity is disrupted frequently .

It is assumed that users are equipped with multi-technology enabled devices. The wireless network environment considered is populated with NetInf-Mobile Nodes (NMN) and Non-NetInf-Mobile Nodes (NNMN). CCU collaborates with NMN & NNMN and provides:

- Probability of NMN & NNMN stay duration in the cellular network

- Cross layer mobility management, predicting the threshold value of the Received Signal Strength RSS to initiate the handover through VNL coordination

- Mobility of nodes through prediction techniques

- Allocation of mobility points/zones and their management

VNL consists of three basic modules which are:

**Handover (HO) module** The Smooth Handover module acts in the scenarios where handover event happens during an ongoing communication session between nodes.

**Power Management Module** The power management module works in the conditions where the mobile node is inactive or idle or left woth low battery power. In such cases, the regular query to update the mobile node location must be suspended. This is very useful in terms of power management of mobile nodes and reducing the signaling overhead of the overall network.

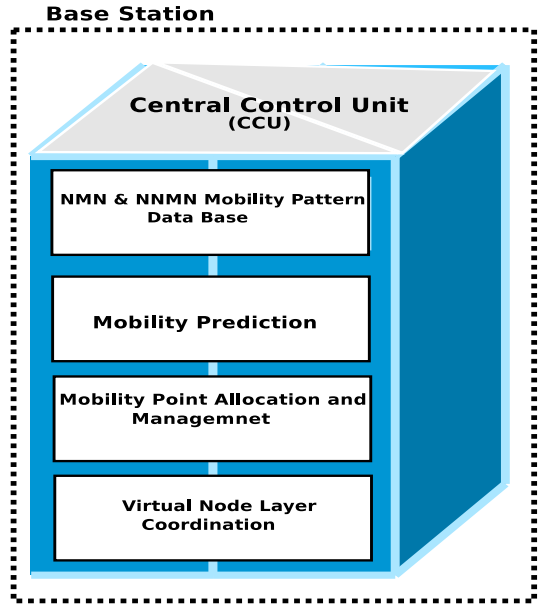

Figure 2: Central Control Unit

**Data Relay Module** The Delay Relay Unit in VNL provides support in challenged environments where there is a frequent disruption in connectivity. It works on the principles of Delay Tolerant Networks (DTN) [?]. The Data Relay Unit is basically designed for two purposes: (i) for different networks interoperability and (ii) to relay or to store the messages in case of disruption in the connectivity. This second feature of relaying and storing messages is what this module does in coordination with ILCTR and OLCTR routing functions.

## 2.2 Cross Layer Support

Cross layer support for mobility management is not a new idea. Here in this case, this support is provided by link and network layers. These two layers measure paramerters like mobile terminal speed, RSS signal and handoff signal delay. As far as handoff signalling delay is concerned, it is measured by sending invalid authenticated messages to the neighbouring base stations and measuring the difference on receiving the reply as suggested in [8]. However, they mentioned an overall increase in the signal headover. In our case, it is done once for each mobile terminal when it is roaming in the Virtual Mobile Point (VMP). CCU updates its record of the mobility pattern. An average value of this delay is taken for each VMP and is used later on. This done for each neighbouring base station. These values are fed into VNL's handoff module.

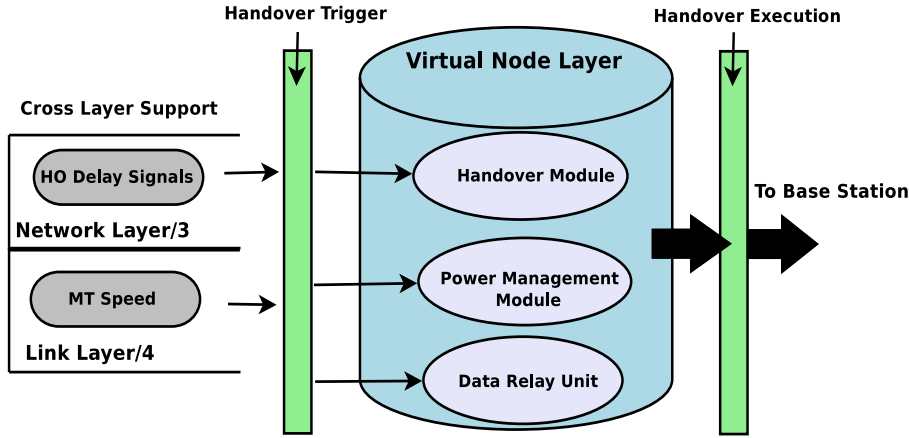

Figure 3: Virtual Node Layer with Cross Layer Support

## 3 Virtual Node Layer Mobility Management

### 3.1 VNL Working Principle

VNL is a prgramming abstraction, implemented on NetInf-MN (NMN). For Non-NetInf-MN (NNMN), Mobile Agent (MA) is used which replicates NMN (NetInf Mobile Node) state on NNMN. Mobile Agent is a computer program or a software that can migrate from one computer (suspending its exucution) to another (resuming again from where it was suspended). The important features of Mobile Agents are autonomy, learning and especially mobility. This last feature makes this technology favorable for the case presented in this work. As they are autonomous in nature, they can migrate to any another computer in the middle of their execution. For distributed application, they are considered to be very powerful tools. They are easily portable and does not require specific system requirement.

### 3.2 Handover Scenario

Before explaining the handover scenario in the context of NetInf, consider the following assumptions:

- The environment is a hetergeneous wireless network environment populated with NMN and NNMN nodes.

- CCU assigns Virtual Mobile Point (VMP) close to the cell boundary of a cellular network. Each VMP has a geographic position with (x,y) coordinates and a span of radius r.

- Each VMP has a center to center distance of r1 from its CCU (Base Statio). The radius of the cell is R.

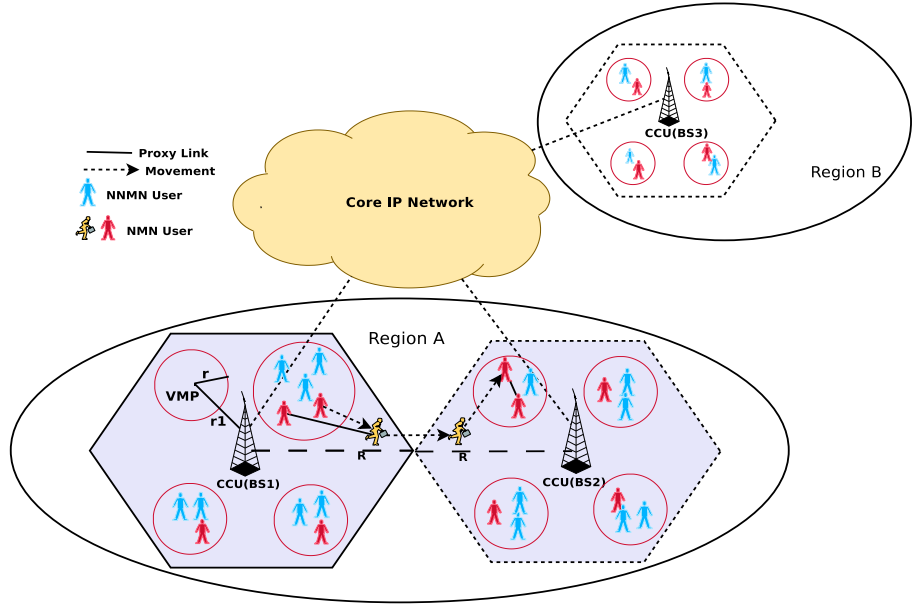

Figure 4: Handover Scenario

- The minimum threshold RSS values within VMP is $\text{VRSS}_{min}$ and within cell its $\text{RSS}_{min}$

- All the nodes in a cell, register their presence. CCU maintains the record and history of each node visiting the cell.

The mobility event occurs in one of the cellular network base station coverage area of geographical Region A, populated with NMN as well as NNMN. In an event a NMN which is in a video conferencing with another user located in Region B as shown in 4, starts moving away from its base station. Mean while, it also moves closer to the boundary of VMP. At base station, with the coordination of CCU, the values of VRSS and RSS are monitored. If the monitored value of the VRSS value less than $\text{VRSS}_{min}$ indicates that it is out of VMP range. At this moment, CCU triggers NMN to make use of the VNL and its modules. In Fig.4, a NMN user while moving away from its base station and out of VMP, activates its handover module. The MA programmed for this task establishes a connection to its nearest NMN to act as a temporary server or data relay node. We name it proxy NMN. The Data Relay Module of this proxy NMN is activated upon receiving the request to perform a task from one of its neighbouring node. Once the proxy node validates NMN identity with the help of CCU, the connection is established. The mobile terminals considered here are euipped with multiple access interface technology. CCU in the mean time is monitoring RSS. Once the measuremnet indicates that the value of RSS is less than $\text{RSS}_{min}$, Base Statio 1 sets free NMN. Now the proxy NMN is relaying data for this node, acting as a temporary server. On the other side, NMN which is now receiving data from the proxy NMN, gets nearer to cell boundary of the neighbouring Base Station 2 as shown. Initially, the RSS signal strength, measured by Base Station 2 is very weak. NMN queries for nearby NMN nodes to establish an adhoc link. On discovering one as shown, it establishes a new proxy link while terminating the old one. As soon as it reaches in the VMP range of Base Station 2, it registers and reconnects with the Base Station 2 terminating its link with the NMN. In this whole process, the video conference session is kept alive without any distortion, sustaining the QoS.

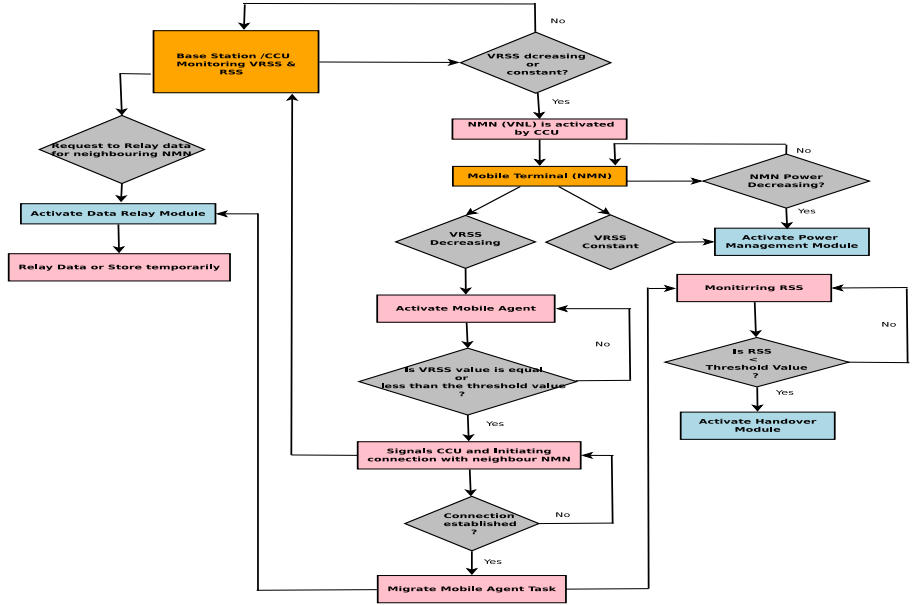

Figure 5: Flow Chart of cross layer supported handover of Network of Information Mobile Node

## 4 Mobility Support in Disconnected Networks

We consider an environment where we have an edge network with nodes having limited or intermittent connectivity [?]. One scenario discussed in [1] is about intermittent connectivity issue in distant areas. The possible solution can be offered based on Delay Tolerant Network (DTN). In NetInf MN architecture, ILCTR and OLCTR under Data Relay Module of VNL, provide routing between NMN and NNMN nodes as well as NMN nodes can act as persistent storage devices in challenged environments where connectivity with the network is disrupted for longer durations. Since NetInf MN is backward compatible with TCP/IP, cases where there is communication between NetInf and Non-NetInf sites or nodes can be carried out with the help of ILCTR, OLCTR and CCU.

# 5 Conclusion & Future Work

We present in this work a handover scenario in a heterogeneous wireless network. The proposed algorithm avoids degradation of QoS during handover and this support is provided by NetInf Mobile Node modules and CCU. The VNL not only leverages QoS but also optimize power consumption and increase persistent storage capacity of the network.

We are in the phase of simulating the handover scenario presented in Fig.4. The idea is to perform our experiments in a heterogeneous wireless network environment. The possible candidates are WiMax and WLAN. Thr future work also includes emulate the whole scenario. The step wise implementation of all modules presented here is in the plan.The final goal is to test the full fledge implementation of the concept.